\documentclass[conference]{IEEEtran}

\usepackage{amsmath,amssymb,amsfonts}
\usepackage{bm}
\usepackage{graphicx}
\usepackage{cite}
\usepackage{tikz}
\usetikzlibrary{positioning, arrows.meta, calc}
\usepackage{url}

\newcommand{\vx}{\bm{x}}
\newcommand{\vy}{\bm{y}}
\newcommand{\vw}{\bm{w}}
\newcommand{\vz}{\bm{z}}
\newcommand{\vr}{\bm{r}}

\newcommand{\mH}{\bm{H}}
\newcommand{\mI}{\bm{I}}
\newcommand{\mSigma}{\bm{\Sigma}}
\newcommand{\R}{\mathbb{R}}
\newcommand{\E}{\mathbb{E}}
\newcommand{\N}{\mathcal{N}}
\DeclareMathOperator{\Tr}{Tr}

\begin{document}

\title{Three-Module SC-VAMP for LDPC-Coded Nonlinear Channels}

\author{
\IEEEauthorblockN{Tadashi Wadayama\IEEEauthorrefmark{1} and Takumi Takahashi\IEEEauthorrefmark{2}}
\IEEEauthorblockA{\IEEEauthorrefmark{1}Nagoya Institute of Technology, Nagoya, Japan\\
Email: wadayama@nitech.ac.jp}
\IEEEauthorblockA{\IEEEauthorrefmark{2}The University of Osaka, Suita, Japan\\
Email: takahashi@comm.eng.osaka-u.ac.jp}
}

\maketitle

\begin{abstract}
We propose a three-module extension of score-based VAMP (SC-VAMP) for signal recovery in nonlinear channels, where the received signal is obtained by applying a nonlinearity to a linear mixture of the transmitted signal, followed by additive Gaussian noise.
The key idea is to introduce a latent variable representing the output of the linear mixing stage, which decomposes the inference problem into three modules: a likelihood module that handles the nonlinear observation via Gauss--Hermite quadrature, a coupling module that enforces the linear constraint between the transmitted signal and the latent variable via LMMSE estimation, and a denoiser module that incorporates the code constraint using belief propagation (BP) decoding.
Each module exchanges extrinsic scalar-Gaussian messages with Onsager corrections derived from posterior variances that are computed in closed form or to quadrature accuracy.
Numerical experiments with BPSK-modulated LDPC codewords transmitted through a hyperbolic tangent channel demonstrate that the proposed method achieves a clear waterfall in bit error rate (BER), with the gap to the capacity estimate narrowing as the block length increases from 128 to 2304.
The framework provides a modular receiver architecture applicable to a broad class of nonlinear channels. Since only the likelihood module depends on the channel nonlinearity, the architecture readily adapts to other channel models by replacing a single module while leaving the coupling and decoder modules unchanged.
\end{abstract}


\section{Introduction}\label{sec:intro}

High-dimensional signal recovery from noisy observations is a fundamental problem in modern communication systems.
In coded systems, the transmitter sends a codeword $\vx$ drawn from a channel code, and the receiver observes $\vy = \mH\vx + \vz$ through a random mixing matrix $\mH$ and additive noise $\vz$.
Approximate message passing (AMP)~\cite{donoho2009amp} and its rotationally invariant extensions, vector AMP (VAMP)~\cite{rangan2019vamp} and orthogonal AMP (OAMP)~\cite{ma2017oamp}, provide powerful iterative frameworks for such linear inverse problems.
Under appropriate random matrix conditions, these algorithms achieve Bayes-optimal performance and admit rigorous state evolution analyses~\cite{bayati2011amp,takeuchi2020vamp,guo2005mutual}.

In practice, however, nonlinearities are often present in the signal path of wireless and optical fiber communication systems.
Power amplifier saturation, analog-to-digital converter (ADC) clipping, and fiber-optic Kerr effects all introduce nonlinear distortions.
The resulting observation model becomes $\vy = f(\mH\vx) + \vz$, where $f(\cdot)$ represents a nonlinear mapping.
Standard VAMP, designed for linear observations, suffers severe model mismatch when applied directly to such channels.
Generalized AMP (GAMP)~\cite{rangan2011gamp} and its VAMP extension (GLM-VAMP)~\cite{schniter2016vamp} decompose the GLM into a nonlinear output channel and a linear mixing stage.
While the channel model $\vy = f(\mH\vx) + \vz$ falls within this GLM class, the algorithm proposed in this paper is not a direct instantiation of GLM-VAMP.
Rather, it reformulates the inference as a compositional SC-VAMP architecture with a uniform score/Fisher-based mean--variance interface that accommodates a BP-based coded prior module within the same extrinsic message-passing protocol.
Specifically, every module represents its posterior mean via Tweedie's formula, derives its Onsager correction from Fisher information or variance ratios, and exchanges scalar-Gaussian extrinsic messages---differing from the Jacobian/divergence-based updates of conventional GLM-VAMP.

In~\cite{wadayama2025scvamp}, we introduced score-based VAMP (SC-VAMP), whose key architectural feature is a uniform mean--variance extrinsic interface shared by all modules.
Every module---whether it handles a linear operation, a nonlinear observation, or a code constraint---receives a Gaussian pseudo-observation parameterized by a mean--variance pair and produces an extrinsic output through a universal Onsager correction formula.
This makes the framework inherently compositional: heterogeneous modules can be freely connected without redesigning existing ones, and extension from two modules to an arbitrary number is straightforward.
Technically, SC-VAMP computes the posterior mean via Tweedie's formula~\cite{efron2011tweedie} and derives the Onsager correction from score functions and Fisher information, enabling a Jacobian-free implementation that also supports data-driven score learning.

The compositionality of SC-VAMP becomes essential when the forward model has multiple distinct components.
The 2-module SC-VAMP in~\cite{wadayama2025scvamp} treats the observation model as $\vy = f(\vx) + \vz$ directly, requiring a single module to handle the entire observation process.
When the forward model involves both a linear mixing $\mH$ and a nonlinearity $f(\cdot)$, as in $\vy = f(\mH\vx) + \vz$, a single observation module cannot cleanly separate these two operations.
A natural solution is to introduce a latent variable $W = \mH X$ and consider the Markov chain $X \to W \to Y$, which factors the joint density into three components: the prior $p_X$, the coupling $p_{W|X}$, and the likelihood $p_{Y|W}$.
Exploiting the compositional architecture of SC-VAMP, we assign one module to each factor: Module~A estimates $W$ from $Y$ based on $p_{Y|W}$, Module~B estimates $X$ based on $p_X$, and Module~C couples $X$ and $W$ based on $p_{W|X}$.
All three modules follow the SC-VAMP extrinsic message-passing protocol: the likelihood and coupling modules compute posterior moments exactly (or to quadrature accuracy) and project them onto scalar-Gaussian messages, while the BP-based denoiser module uses approximate posterior moments.
More broadly, the proposed architecture is not tied to a particular nonlinearity; by replacing Module~A with an appropriate score estimator, the same framework extends naturally to channels of the form $\vy = f(\mH\vx) + \vz$.

The proposed framework can also be compared with \emph{turbo equalization}~\cite{douillard1995turbo,tuchler2002turbo} and its extensions to MIMO detection~\cite{wang1999iterative}, which iterate between a soft equalizer and a soft channel decoder, exchanging extrinsic LLRs.
These methods, however, are typically designed for specific linear channel structures, and the heuristic LLR subtraction used to compute extrinsic information lacks a principled justification, whereas SC-VAMP derives extrinsic messages from the Onsager correction grounded in score functions and Fisher information.

Our contributions are as follows:
\begin{enumerate}
  \item We extend the score/Fisher parameterization of SC-VAMP~\cite{wadayama2025scvamp} to a three-module architecture, connecting a nonlinear likelihood module (Gauss--Hermite quadrature~\cite{abramowitz1964handbook}), a linear coupling module (LMMSE), and a coded prior module (BP decoding~\cite{gallager1962ldpc,richardson2008modern}) through a uniform mean--variance extrinsic interface.
  \item We incorporate an LDPC BP decoder as a symbol-domain SISO denoiser within the same Onsager-corrected extrinsic protocol, providing a principled alternative to the heuristic LLR subtraction used in classical turbo receivers.
  \item We demonstrate on the $f = \tanh$ channel that the method achieves a clear waterfall BER curve sharpening from $N = 128$ to $2304$, with a $\sim$3~dB gain from the Onsager correction.
\end{enumerate}

\section{Preliminaries}\label{sec:prelim}

\subsection{Notation}
Vectors are denoted by bold lowercase letters (e.g., $\vx$) and matrices by bold uppercase letters (e.g., $\mH$).
Random variables (and random vectors) are denoted by uppercase letters ($X$, $Y$), and their realizations by lowercase letters ($x$, $y$, or bold $\vx$, $\vy$ for vectors).
We write $\N(\bm{\mu}, \mSigma)$ for a Gaussian distribution with mean $\bm{\mu}$ and covariance $\mSigma$, and $\N(x; \mu, \sigma^2)$ for the corresponding density evaluated at $x$.
For a continuous random vector $U \in \R^n$ with density $p_U$, the \emph{score function} is $s_U(\bm{u}) \triangleq \nabla_{\bm{u}} \log p_U(\bm{u})$ and the \emph{scalar Fisher information} is $J_U \triangleq \E\bigl[\|s_U(U)\|^2\bigr]$.
The conditional counterparts are $s_{U|Y}(\bm{u}|\vy) \triangleq \nabla_{\bm{u}} \log p_{U|Y}(\bm{u}|\vy)$ and $J_{U|Y} \triangleq \E\bigl[\|s_{U|Y}(U|Y)\|^2\bigr]$.
Throughout, score functions are applied to pseudo-observations $R = X + \sqrt{v}\,N$, which have smooth densities via Gaussian convolution even when $X$ is discrete.

\subsection{System Model}\label{sec:system}

Consider a coded system in which the transmitter encodes $K$ information bits into an LDPC codeword, maps it to BPSK, producing a random vector $X \in \{+1, -1\}^N$.
The received signal is
\begin{equation}\label{eq:channel}
  Y = f(\mH X) + Z,
\end{equation}
where $\mH \in \R^{M \times N}$ is the channel matrix with i.i.d.\ entries $H_{ij} \sim \N(0, 1/M)$, $f\colon \R^M \to \R^M$ is a nonlinear mapping, and $Z \sim \N(\bm{0}, \sigma^2 \mI_M)$ is additive white Gaussian noise.
The signal-to-noise ratio is defined as $\mathrm{SNR} = 1/\sigma^2$.

We introduce the latent random vector $W \triangleq \mH X \in \R^M$, so that the observation becomes $Y = f(W) + Z$.
This yields the Markov chain $X \to W \to Y$ and the joint density
\begin{equation}\label{eq:factor}
  p(\vx, \vw, \vy) = p_X(\vx) \cdot \delta(\vw - \mH\vx) \cdot p_{Y|W}(\vy|\vw),
\end{equation}
where $\vx, \vw, \vy$ denote realizations of $X, W, Y$, respectively, $p_{Y|W}(\vy|\vw) = \N(\vy; f(\vw), \sigma^2\mI_M)$, and $\delta(\cdot)$ denotes the Dirac delta function.
The three factors on the right-hand side correspond to the code constraint, the linear coupling, and the nonlinear likelihood, respectively.
This factorization is the basis of the 3-module decomposition proposed in Section~\ref{sec:method}.

\subsection{Review of 2-Module SC-VAMP}\label{sec:review}

The SC-VAMP framework proposed in~\cite{wadayama2025scvamp} operates by iterating two SISO (soft-input soft-output) modules that exchange extrinsic mean--variance messages.
Each module acts as an estimator: given an input pair $(\vr_{in}, v_{in})$ representing a Gaussian pseudo-observation $\vr_{in} = \vx + \vz$, $\vz \sim \N(\bm{0}, v_{in}\mI)$, it computes a posterior mean $\vx_{post}$ and produces an extrinsic output $(\vx_{out}, v_{out})$.

The posterior mean is computed via Tweedie's formula~\cite{efron2011tweedie}:
\begin{equation}\label{eq:tweedie}
  \vx_{post} = \vr_{in} + v_{in}\, s_{R_{in}|Y}(\vr_{in}|\vy),
\end{equation}
where $s_{R_{in}|Y}$ is the conditional score function of the pseudo-observation $R_{in}$ given $Y$.
The posterior variance is $v_{post} = v_{in}\,\alpha(v_{in})$, where $\alpha(v_{in})$ is the Onsager coefficient.
The extrinsic output is obtained by ``subtracting'' the input information:
\begin{equation}\label{eq:ext_mean}
  \vx_{out} = \frac{\vx_{post} - \alpha\, \vr_{in}}{1 - \alpha}, \qquad
  v_{out} = \frac{\alpha}{1 - \alpha}\, v_{in},
\end{equation}
with
\begin{equation}\label{eq:alpha_def}
  \alpha = \frac{v_{post}}{v_{in}} = 1 - \frac{v_{in}}{N}\,J_{R_{in}|Y},
\end{equation}
where $J_{R_{in}|Y} = \E\bigl[\|s_{R_{in}|Y}(R_{in}|Y)\|^2\bigr]$ is the scalar (trace) Fisher information.
The Onsager coefficient $\alpha$ prevents ``double-counting'' of information in loopy message passing and is expressed directly through $J_{R_{in}|Y}$, enabling a Jacobian-free implementation.

\section{Proposed Method: Multi-Module SC-VAMP}\label{sec:method}

\subsection{Generic Framework}\label{sec:decomp}

The multi-module SC-VAMP framework naturally extends to general Markov chains $X_1 \to X_2 \to \cdots \to X_L$ by introducing a coupling module for each edge $(X_{k-1}, X_k)$, with the two end modules corresponding to the prior and likelihood factors.
For simplicity of presentation, we focus on the 3-module case with a single latent variable, i.e., the chain $X \to W \to Y$, where $X$ is the signal to be estimated and $Y$ is the observed signal.

\begin{figure}[t]
\centering
\begin{tikzpicture}[
  block/.style={draw, thick, minimum width=2.0cm, minimum height=1.2cm, align=center, font=\small},
  arr/.style={-{Stealth[length=2.5mm]}, thick},
  lbl/.style={font=\scriptsize, fill=white, inner sep=1pt}
]
\node[block] (C) at (0, 0) {Module~C\\(Coupling)\\$p_{W|X}$};
\node[block] (B) at (4.5, 0) {Module~B\\(Prior)\\$p_X$};
\node[block] (A) at (0, -2.8) {Module~A\\(Likelihood)\\$p_{Y|W}$};

\draw[arr] ([yshift=3pt]C.east) -- ([yshift=3pt]B.west)
  node[lbl, midway, above] {$(\vr_x^{C,out}, v_x^{C,out})$};
\draw[arr] ([yshift=-3pt]B.west) -- ([yshift=-3pt]C.east)
  node[lbl, midway, below] {$(\vr_x^{B,out}, v_x^{B,out})$};

\draw[arr] ([xshift=-3pt]C.south) -- ([xshift=-3pt]A.north)
  node[lbl, midway, left] {$(\vr_w^{C,out}, v_w^{C,out})$};
\draw[arr] ([xshift=3pt]A.north) -- ([xshift=3pt]C.south)
  node[lbl, midway, right] {$(\vr_w^{A,out}, v_w^{A,out})$};

\draw[arr] (-1.8, -2.8) -- (A.west)
  node[lbl, midway, above] {$\vy$};

\end{tikzpicture}
\caption{Block diagram of 3-module SC-VAMP. Module~C (coupling, $p_{W|X}$) exchanges extrinsic messages with Module~B (prior, $p_X$) on the $X$-side and with Module~A (likelihood, $p_{Y|W}$) on the $W$-side. Each arrow carries a mean--variance pair.}
\label{fig:block}
\end{figure}
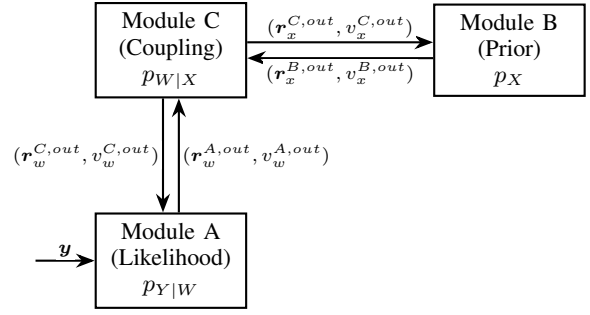

The factorization~\eqref{eq:factor} in Section~\ref{sec:system} naturally suggests three SISO modules, each responsible for one factor:
\begin{itemize}
  \item \textbf{Module~A} (Likelihood): estimates $W$ from $p_{Y|W}$ and the observation $\vy$.
  \item \textbf{Module~B} (Prior/Denoiser): estimates $X$ from $p_X$.
  \item \textbf{Module~C} (Coupling): estimates $(X, W)$ jointly from $p_{W|X}$.
\end{itemize}
Module~C acts as a central hub, exchanging extrinsic messages with Module~A on the $W$-side and with Module~B on the $X$-side (Fig.~\ref{fig:block}).
Each module refines its estimate by combining the pseudo-observation with its own factor of the joint density, and the Onsager correction ensures that the extrinsic output does not ``double-count'' the input information.

The proposed 3-module SC-VAMP rests on the following two assumptions.

\begin{enumerate}
\item \textbf{Assumption~1 (Markov chain).}
  The random variables $X$, $W$, $Y$ form a Markov chain $X \to W \to Y$, i.e., the joint density factorizes as
  \begin{equation}\label{eq:markov}
    p_{XWY}(\vx, \vw, \vy) = p_X(\vx)\, p_{W|X}(\vw|\vx)\, p_{Y|W}(\vy|\vw),
  \end{equation}
  where $p_{W|X}$ is a general stochastic kernel (deterministic maps such as $W = f_1(X)$ are included as special cases via a Dirac delta).

\item \textbf{Assumption~2 (Gaussian pseudo-observation approximation).}
  All inter-module messages can be approximated as Gaussian pseudo-observations parameterized by a mean--variance pair $(\vr, v)$.
  Specifically, Module~B sends $(\vr_x, v_x)$ to Module~C and Module~A sends $(\vr_w, v_w)$ to Module~C, each interpreted as
  \begin{equation}\label{eq:pseudo}
    R_X = X + \sqrt{v_x}\, N_X, \qquad R_W = W + \sqrt{v_w}\, N_W,
  \end{equation}
  where $N_X \sim \N(\bm{0}, \mI_N)$ and $N_W \sim \N(\bm{0}, \mI_M)$ are independent of each other and of $(X, W)$.
  This is the same Gaussian message assumption employed in 2-module SC-VAMP~\cite{wadayama2025scvamp}.
\end{enumerate}

\subsection{Module Definitions (Score-Based)}\label{sec:generic_modules}

Each module computes a posterior mean via Tweedie's formula and produces an extrinsic output with Onsager correction, following the SC-VAMP protocol~\eqref{eq:tweedie}--\eqref{eq:ext_mean}.
The uniform interface described in Section~\ref{sec:intro} allows heterogeneous modules to be composed without modification.

\subsubsection{Module~C (Coupling)}
Given inputs $(\vr_x, v_x)$ from Module~B and $(\vr_w, v_w)$ from Module~A, Module~C incorporates solely its own factor $p_{W|X}$ by forming the \emph{tilted distribution}
\begin{align}\label{eq:tilted_C}
  \tilde{p}^{C}(\vx, \vw) \triangleq \frac{1}{Z_C}\, &p_{W|X}(\vw|\vx) \nonumber \\
  &\cdot \N(\vx; \vr_x, v_x \mI_N)\, \N(\vw; \vr_w, v_w \mI_M),
\end{align}
where $Z_C(\vr_x, \vr_w)$ is the normalization constant obtained by integrating out $(\vx, \vw)$.
The two Gaussian factors are the incoming messages from Modules~B and~A (Assumption~2), not the marginal priors $p_X$ or $p_W$.
The posterior means equal
\begin{align}
  \hat{\vx}_{post}^{C} &= \E_{\tilde{p}^C}[\vx] = \vr_x + v_x\, \nabla_{\vr_x} \log Z_C, \label{eq:tweedie_C_x} \\
  \hat{\vw}_{post}^{C} &= \E_{\tilde{p}^C}[\vw] = \vr_w + v_w\, \nabla_{\vr_w} \log Z_C, \label{eq:tweedie_C_w}
\end{align}
where $\nabla_{\vr_x} \log Z_C$ and $\nabla_{\vr_w} \log Z_C$ are the score functions of the tilted distribution~\eqref{eq:tilted_C} with respect to $\vr_x$ and $\vr_w$, respectively (Tweedie's formula applied to each side).
The posterior variances and Onsager coefficients are
\begin{equation}\label{eq:alpha_C}
  v_{post,x}^{C} = \frac{1}{N}\Tr\!\bigl(\mathrm{Cov}_{\tilde{p}^C}(\vx)\bigr),\quad
  \alpha_x^{C} = \frac{v_{post,x}^{C}}{v_x},
\end{equation}
and similarly $v_{post,w}^{C} = (1/M)\,\Tr\!\bigl(\mathrm{Cov}_{\tilde{p}^C}(\vw)\bigr)$, $\alpha_w^{C} = v_{post,w}^{C}/v_w$.
The variance-ratio form $\alpha = v_{post}/v_{in}$ is equivalent to the Fisher-information form~\eqref{eq:alpha_def} via the identity $v_{post}/v_{in} = 1 - (v_{in}/N)\,J$; here we use the variance-ratio form because the posterior covariance of Module~C is available in closed form (see Section~\ref{sec:modC}).
The extrinsic outputs on each side are
\begin{align}
  \vr_x^{C,out} &= \frac{\hat{\vx}_{post}^{C} - \alpha_x^{C}\, \vr_x}{1 - \alpha_x^{C}}, &
  v_x^{C,out} &= \frac{\alpha_x^{C}}{1 - \alpha_x^{C}}\, v_x, \label{eq:ext_C_x} \\
  \vr_w^{C,out} &= \frac{\hat{\vw}_{post}^{C} - \alpha_w^{C}\, \vr_w}{1 - \alpha_w^{C}}, &
  v_w^{C,out} &= \frac{\alpha_w^{C}}{1 - \alpha_w^{C}}\, v_w. \label{eq:ext_C_w}
\end{align}
This procedure---tilted distribution, moment matching, extrinsic output---is equivalent to the EP projection step~\cite{minka2001ep,opper2005ep}, with the incoming messages as the cavity distribution and the extrinsic output as the updated site approximation.

\subsubsection{Module~A (Likelihood)}
Given input $(\vr_w, v_w)$ from Module~C and observation $\vy$, Module~A forms the tilted distribution $\tilde{p}^{A}(\vw) \propto p_{Y|W}(\vy | \vw)\, \N(\vw; \vr_w, v_w \mI_M)$ with normalization constant $Z_A(\vr_w, \vy) \triangleq \int p_{Y|W}(\vy | \vw)\, \N(\vw; \vr_w, v_w \mI_M)\, d\vw$.
The posterior mean is computed via Tweedie's formula:
\begin{equation}\label{eq:tweedie_A}
  \hat{\vw}_{post}^{A} = \vr_w + v_w\, \bm{s}_{R_W|Y}^{(A)}(\vr_w \mid \vy),
\end{equation}
where $\bm{s}_{R_W|Y}^{(A)}(\vr_w \mid \vy) \triangleq \nabla_{\vr_w} \log Z_A(\vr_w, \vy)$ is the score of the tilted distribution.
The Onsager coefficient is $\alpha^{A} = v_{post}^{A}/v_w$ with $v_{post}^{A} = (1/M)\,\Tr\bigl(\mathrm{Cov}_{\tilde{p}^A}(\vw)\bigr)$, and the extrinsic output follows~\eqref{eq:ext_mean}.

\subsubsection{Module~B (Prior/Denoiser)}
Given input $(\vr_x, v_x)$ from Module~C, Module~B computes
\begin{equation}\label{eq:tweedie_B}
  \hat{\vx}_{post}^{B} = \vr_x + v_x\, \bm{s}_{R_X}^{(B)}(\vr_x),
\end{equation}
where $\bm{s}_{R_X}^{(B)}(\vr_x) \triangleq \nabla_{\vr_x} \log p_{R_X}(\vr_x)$ is the unconditional score with $p_{R_X}(\vr_x) = \int p_X(\vx)\, \N(\vr_x; \vx, v_x \mI_N)\, d\vx$.
Module~B has no dependence on $\vy$ and acts purely as a denoiser based on the prior $p_X$.
In practice (e.g., BP decoding), the Onsager coefficient is approximated by $\alpha^{B} = v_{post}^{B}/v_x$ (see Section~\ref{sec:modB}).
The extrinsic output follows~\eqref{eq:ext_mean}.

\subsubsection{Iteration Schedule}
One iteration consists of: (i)~Module~C produces extrinsic outputs to both sides; (ii)~Module~B updates $(\vr_x, v_x)$; (iii)~Module~A updates $(\vr_w, v_w)$; and these feed back into Module~C.
Steps (ii) and (iii) are independent and may be executed in parallel.

\subsection{Instantiation for Nonlinear Coded Channels}\label{sec:instance}

We now instantiate the generic framework for the nonlinear channel model $\vy = f(\mH\vx) + \vz$ introduced in Section~\ref{sec:system}.
By setting the latent variable $W = \mH X$, the Markov chain $X \to W \to Y$ yields the three factors in~\eqref{eq:factor}: the prior $p_X$ (LDPC code constraint), the coupling $p_{W|X}(\vw|\vx) = \delta(\vw - \mH\vx)$ (deterministic linear map), and the likelihood $p_{Y|W}(\vy|\vw) = \N(\vy; f(\vw), \sigma^2\mI_M)$ (nonlinear observation with AWGN).
For each module, we derive concrete computations that replace the abstract score functions in~\eqref{eq:tweedie_C_x}--\eqref{eq:tweedie_B}.
The iteration follows the schedule in Section~\ref{sec:generic_modules}, initialized with $(\vr_x, v_x) = (\bm{0}, 1)$ and $(\vr_w, v_w) = (\vy, \sigma^2)$.
The $X$-side initialization corresponds to a non-informative prior (zero mean, unit variance for BPSK).
The $W$-side initialization $(\vr_w, v_w) = (\vy, \sigma^2)$ is exact when $f$ is the identity, and is a heuristic for general $f$ that only affects the first iteration, as Module~A replaces it with a properly calibrated message from the second iteration onward.
After $T$ iterations, the hard decision is $\hat{x}_i = \mathrm{sgn}(\hat{x}_{post,i}^{B})$.

\subsubsection{Module~C: SISO-LMMSE}\label{sec:modC}

Since $W = \mH X$ is a deterministic linear map, the coupling $p_{W|X}(\vw|\vx) = \delta(\vw - \mH\vx)$ eliminates $\vw$ as an independent variable.
Substituting into the pseudo-observation model~\eqref{eq:pseudo}, we have $R_W = \mH X + \sqrt{v_w}\, N_W$ in addition to $R_X = X + \sqrt{v_x}\, N_X$.
Since both observations are linear in $X$ with Gaussian noise, the posterior of $X$ given $(R_X, R_W)$ is Gaussian.
The posterior precision is the sum of the individual precisions, $(1/v_x)\mI_N$ from $R_X$ and $(1/v_w)\mH^\top\mH$ from $R_W$, yielding
\begin{align}
  \hat{\vx}_{post}^{C} &= \mSigma_x^{C} \left(\frac{\vr_x}{v_x} + \frac{\mH^\top \vr_w}{v_w}\right), \label{eq:xpost_C} \\
  \mSigma_x^{C} &= \left(\frac{1}{v_x}\mI_N + \frac{1}{v_w}\mH^\top\mH\right)^{-1}, \label{eq:Sigma_C}
\end{align}
and $\hat{\vw}_{post}^{C} = \mH\, \hat{\vx}_{post}^{C}$.
The scalar posterior variances are
\begin{equation}\label{eq:vpost_C}
  v_{post,x}^{C} = \frac{1}{N}\Tr\bigl(\mSigma_x^{C}\bigr), \quad
  v_{post,w}^{C} = \frac{1}{M}\Tr\bigl(\mH\mSigma_x^{C}\mH^\top\bigr).
\end{equation}
Using the eigendecomposition $\mH^\top\mH = \bm{U}\bm{\Lambda}\bm{U}^\top$ with eigenvalues $\lambda_1, \ldots, \lambda_N$, the traces in~\eqref{eq:vpost_C} and the Onsager coefficients $\alpha_x^{C} = v_{post,x}^{C}/v_x$, $\alpha_w^{C} = v_{post,w}^{C}/v_w$ reduce to sums over eigenvalues, e.g., $\alpha_x^{C} = \frac{1}{N}\sum_{n=1}^{N} \frac{v_w}{v_w + v_x \lambda_n}$.
The extrinsic outputs on the $X$- and $W$-sides then follow the generic formulas~\eqref{eq:ext_C_x}--\eqref{eq:ext_C_w}.

\subsubsection{Module~A: Nonlinear Likelihood}\label{sec:modA}

Module~A receives the extrinsic output $(\vr_w^{C,out}, v_w^{C,out})$ from Module~C on the $W$-side as its input pseudo-observation, together with the channel output $\vy$.
Writing $(\vr_w^{in}, v_w^{in}) = (\vr_w^{C,out}, v_w^{C,out})$ for brevity, Module~A estimates $W$ by incorporating the likelihood $p_{Y|W}(\vy|\vw) = \N(\vy; f(\vw), \sigma^2\mI_M)$.
The posterior mean is computed via Tweedie's formula:
\begin{equation}\label{eq:wpost_A}
  \hat{\vw}_{post}^{A} = \vr_w^{in} + v_w^{in}\, \bm{s}_{R_W|Y}^{(A)}(\vr_w^{in} \mid \vy),
\end{equation}
where $\bm{s}_{R_W|Y}^{(A)}(\vr_w^{in} \mid \vy) = \nabla_{\vr_w^{in}} \log Z_A(\vr_w^{in}, \vy)$ is the score of the tilted distribution of Module~A.
The posterior variance is $v_{post}^{A} = (1/M)\,\Tr\bigl(\mathrm{Cov}_{\tilde{p}^A}(\vw)\bigr)$ and the Onsager coefficient is $\alpha^{A} = v_{post}^{A}/v_w^{in}$.
The extrinsic output follows~\eqref{eq:ext_mean} and serves as the updated $(\vr_w, v_w)$ for Module~C in the next iteration.

The score function can be evaluated in various ways depending on the structure of $f$.
When $f$ is applied component-wise, i.e., $[f(\vw)]_m = f(w_m)$ for a scalar function $f\colon \R \to \R$, the score decomposes as $[\bm{s}_{R_W|Y}^{(A)}(\vr_w^{in} \mid \vy)]_m = s_m^{(A)}(r_{w,m}^{in} \mid y_m)$, where each scalar score is given by
\begin{equation}\label{eq:score_A_scalar}
  s_m^{(A)}(r_{w,m}^{in} \mid y_m) = \frac{\E[W_m \mid r_{w,m}^{in}, y_m] - r_{w,m}^{in}}{v_w^{in}}
\end{equation}
with
\begin{align}\label{eq:postmean_A}
  &\E[W_m \mid r_{w,m}^{in}, y_m] \nonumber \\
  &\quad = \frac{\int w\, \N(w; r_{w,m}^{in}, v_w^{in})\, \N(y_m; f(w), \sigma^2)\, dw}{\int \N(w; r_{w,m}^{in}, v_w^{in})\, \N(y_m; f(w), \sigma^2)\, dw}.
\end{align}
The posterior variance is computed from the same integrals:
\begin{equation}\label{eq:vpost_A}
  v_{post}^{A} = \frac{1}{M}\sum_{m=1}^{M}\bigl(\E[W_m^2 \mid r_{w,m}^{in}, y_m] - (\E[W_m \mid r_{w,m}^{in}, y_m])^2\bigr),
\end{equation}
where $\E[W_m^2 \mid r_{w,m}^{in}, y_m]$ is obtained by replacing $w$ with $w^2$ in the numerator of~\eqref{eq:postmean_A}.
Both moments can be efficiently evaluated by Gauss--Hermite quadrature~\cite{abramowitz1964handbook} with a single set of quadrature points.
When $f$ is not component-wise---or when $f$ is unknown---the score can be learned from data (e.g., via flow matching or denoising score matching), making Module~A applicable to general nonlinear channels without requiring an analytical expression for $f$.
When $f$ is the identity, Module~A reduces to a simple linear Gaussian update that returns $(\vr_w^{A,out}, v_w^{A,out}) = (\vy, \sigma^2)$ at every iteration; substituting this into Module~C reduces to the 2-module SC-VAMP formulation of~\cite{wadayama2025scvamp} for the linear channel $\vy = \mH\vx + \vz$.

To verify this claim, note that when $f = \mathrm{id}$, Module~A yields $\alpha^{A} = \sigma^2/(v_w^{in} + \sigma^2)$ and the extrinsic output $(\vr_w^{A,out}, v_w^{A,out}) = (\vy, \sigma^2)$, which is constant across iterations.
Substituting into Module~C's LMMSE equations~\eqref{eq:xpost_C}--\eqref{eq:Sigma_C} and applying the matrix inversion lemma recovers the observation-module update of the 2-module SC-VAMP in~\cite{wadayama2025scvamp}, confirming that the 3-module iteration reduces exactly to the 2-module case.

\subsubsection{Module~B: BP Decoder as Denoiser}\label{sec:modB}

Module~B receives the extrinsic output $(\vr_x^{C,out}, v_x^{C,out})$ from Module~C on the $X$-side as its input pseudo-observation.
Writing $(\vr_x^{in}, v_x^{in}) = (\vr_x^{C,out}, v_x^{C,out})$, Module~B incorporates the LDPC code constraint $p_X$ and acts as a SISO denoiser.
The Gaussian pseudo-observation $r_{x,i}^{in} = x_i + \sqrt{v_x^{in}}\,n_i$ with BPSK $x_i \in \{+1, -1\}$ is converted into input LLRs $L_i^{in} = 2 r_{x,i}^{in}/v_x^{in}$, which represent the channel reliability of each bit.
These LLRs are fed into a BP decoder~\cite{gallager1962ldpc,richardson2008modern} (implemented using the Kaira library~\cite{kaira2025}), which iteratively exchanges messages along the edges of the LDPC Tanner graph and produces a-posteriori LLRs $L_i^{app}$.
The posterior mean for each bit is $\hat{x}_{post,i}^{B} = \tanh(L_i^{app}/2) = \E[X_i | L_i^{app}]$ for BPSK signaling, and the posterior variance is
\begin{equation}\label{eq:bp_post}
  v_{post}^{B} = \frac{1}{N}\sum_{i=1}^{N}\bigl(1 - (\hat{x}_{post,i}^{B})^2\bigr).
\end{equation}
The Onsager coefficient is set to $\alpha^{B} = v_{post}^{B}/v_x^{in}$, and the extrinsic output returned to Module~C is
\begin{equation}\label{eq:ext_B}
  \vr_x^{B,out} = \frac{\hat{\vx}_{post}^{B} - \alpha^{B}\, \vr_x^{in}}{1 - \alpha^{B}}, \qquad
  v_x^{B,out} = \frac{\alpha^{B}}{1 - \alpha^{B}}\, v_x^{in},
\end{equation}
serving as the updated pseudo-observation $(\vr_x, v_x) \leftarrow (\vr_x^{B,out}, v_x^{B,out})$ for Module~C.

Since BP is an approximate inference algorithm on loopy factor graphs, $\hat{\vx}_{post}^{B}$ does not equal the true MMSE estimate $\E[\vx \mid \vr_x^{in}]$ under $p_X$.
The variance-ratio form $\alpha^{B} = v_{post}^{B}/v_x^{in}$ coincides with the theoretically correct Onsager correction (the average divergence of the denoiser) only for the exact MMSE denoiser; for BP, it serves as a variance-ratio surrogate.
This approximation is standard in turbo-type receivers~\cite{tuchler2002turbo,wadayama2025scvamp} and works well in practice.
In implementation, all Onsager coefficients are clipped to $[\epsilon, 1-\epsilon]$ (e.g., $\epsilon = 10^{-6}$) to ensure that the extrinsic variances remain positive and finite.
The extrinsic exchange between Module~B and Module~C structurally corresponds to the decoder--detector interaction in classical iterative detection and decoding (IDD)~\cite{tuchler2002turbo,wang1999iterative}; the SC-VAMP framework subsumes this interaction by replacing heuristic LLR subtraction with the Onsager-corrected mean--variance interface.

\section{Numerical Experiments}\label{sec:exp}

Throughout this section, we use rate-$1/2$ LDPC codes (CCSDS and WiMax) with BPSK modulation, $T = 20$ outer iterations, and $20$ BP iterations per outer iteration.

\subsection{Effect of Onsager Correction}\label{sec:exp_onsager}

We first demonstrate that the Onsager correction is essential for proper convergence of the 3-module SC-VAMP.
To isolate the effect of the correction from the nonlinearity, we set $f = \mathrm{id}$ (i.e., $\vy = \mH\vx + \vz$) and compare three methods:
\begin{itemize}
  \item \textbf{SC-VAMP} (proposed): Full Onsager correction on all modules.
  \item \textbf{No Onsager}: Posteriors are passed directly without extrinsic subtraction.
  \item \textbf{LLR Turbo}: Modules~C and~A use Onsager correction, but Module~B uses classical LLR subtraction $L_\mathrm{ext} = L_\mathrm{app} - L_\mathrm{in}$.
\end{itemize}
For each (SNR, seed) pair, all three methods share the same $\mH$, $\vx$, and $\vz$.

\begin{figure}[t]
  \centering
  \includegraphics[width=\columnwidth]{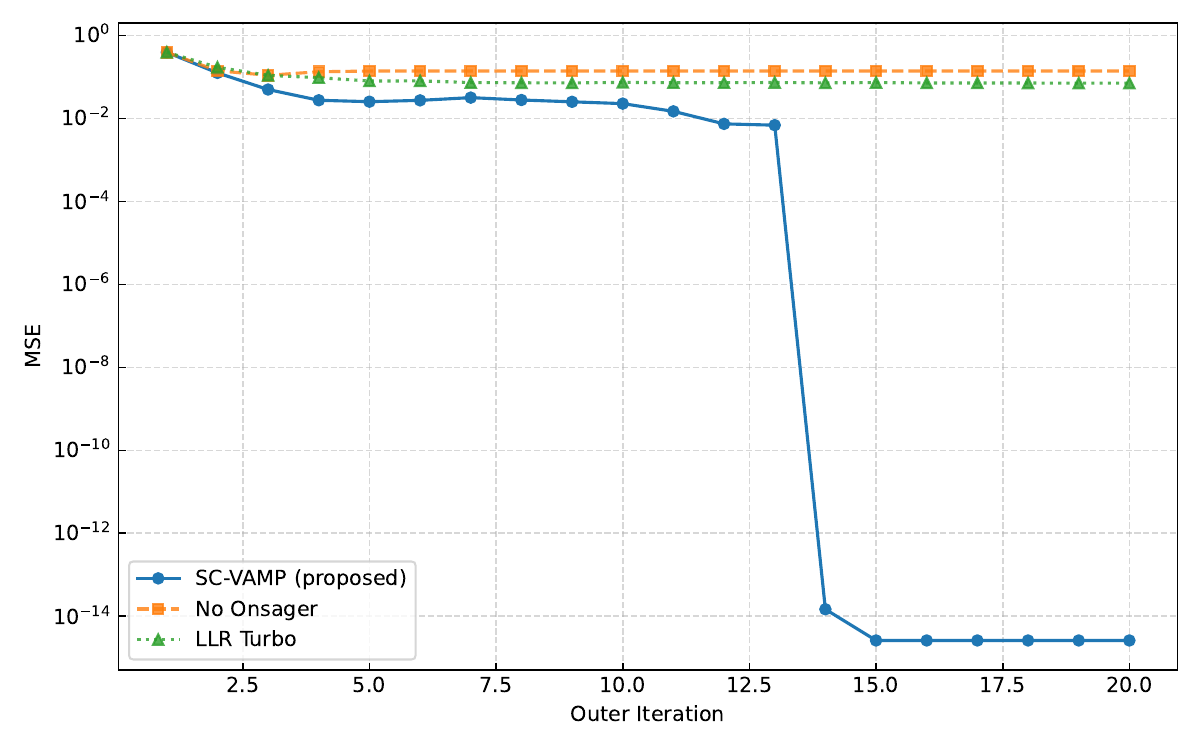}
  \caption{MSE versus outer iteration at $\mathrm{SNR} = 6$~dB ($f = \mathrm{id}$), averaged over 50 trials.
  SC-VAMP converges to machine precision ($\sim 10^{-15}$) at iteration~14,
  while No~Onsager and LLR~Turbo stall at MSE $\approx 10^{-1}$.}
  \label{fig:conv_onsager}
\end{figure}

Fig.~\ref{fig:conv_onsager} shows the MSE $\frac{1}{N}\|\hat{\vx} - \vx\|^2$ versus the outer iteration at $\mathrm{SNR} = 6$~dB, averaged over 50 trials.
SC-VAMP converges to machine precision ($\sim 10^{-15}$) at iteration~14, while No~Onsager and LLR~Turbo stall at MSE $\approx 10^{-1}$ due to double-counting of self-information.

\begin{figure}[t]
  \centering
  \includegraphics[width=\columnwidth]{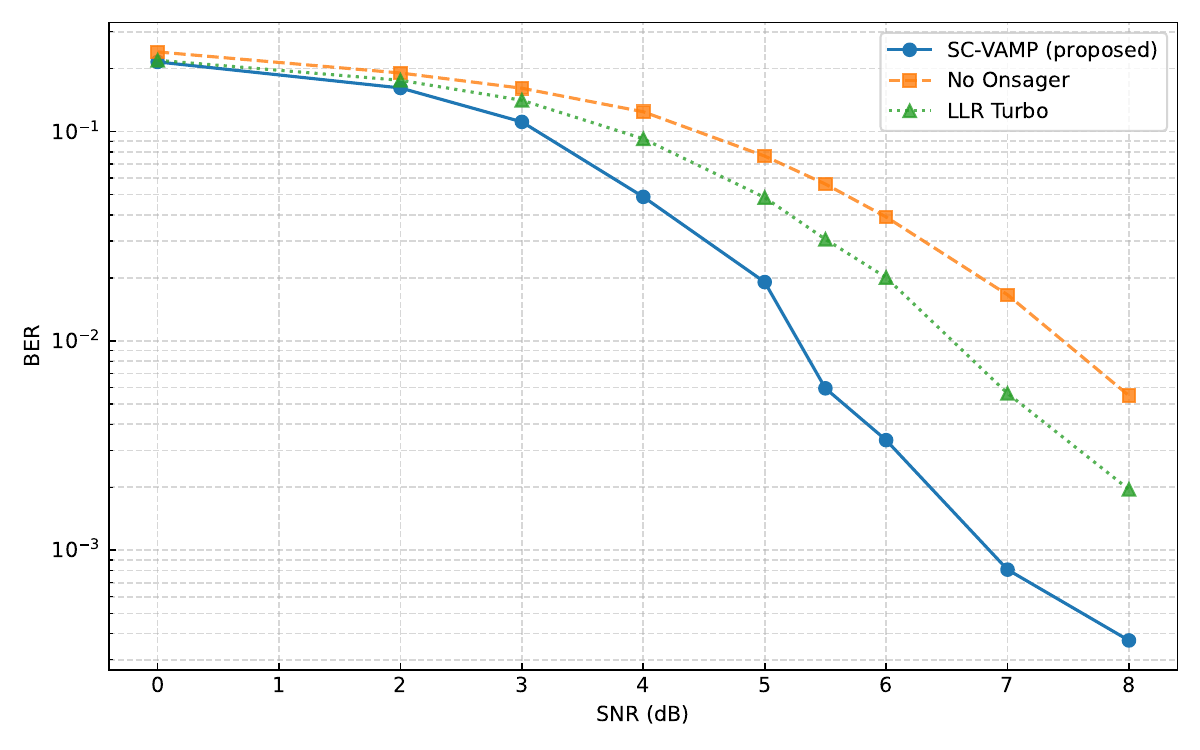}
  \caption{BER versus SNR for $f = \mathrm{id}$ channel with LDPC$(128, 64)$.
  Adaptive seeding (min.\ 500 errors, max 2000 seeds) is used for statistical reliability.
  Only SNR points with $\mathrm{BER} > 0$ are shown.}
  \label{fig:ber_onsager}
\end{figure}

Fig.~\ref{fig:ber_onsager} shows the BER versus SNR (adaptive seeding: min.\ 500 errors or max 2000 seeds per SNR point).
SC-VAMP achieves the steepest waterfall: at $\mathrm{BER} = 10^{-2}$, it requires $\approx$5~dB versus $\approx$8~dB for No~Onsager---a gap of $\sim$3~dB.
The ordering $\mathrm{BER}_\mathrm{SC\text{-}VAMP} \leq \mathrm{BER}_\mathrm{LLR} \leq \mathrm{BER}_\mathrm{No\text{-}Onsager}$ holds at every SNR, confirming that the symbol-domain Onsager correction outperforms classical LLR subtraction.

\subsection{Nonlinear Channel ($f=\tanh$)}\label{sec:exp_tanh}

We now evaluate the 3-module SC-VAMP on the nonlinear channel $\vy = \tanh(\mH\vx) + \vz$, the primary target application.
Module~A handles $f = \tanh$ via Gauss--Hermite quadrature with $Q = 50$ points; all modules use Onsager correction and the same adaptive seeding as Section~\ref{sec:exp_onsager}.
As a reference, a mismatched 2-module SC-VAMP baseline that ignores the nonlinearity (treating the channel as $\vy = \mH\vx + \vz$) is also plotted for $N = 512$.

To isolate coding gain, we use a block-diagonal $\mH = \mathrm{blockdiag}(\mH_b, \ldots, \mH_b)$ with a shared i.i.d.\ Gaussian block $\mH_b \in \R^{32 \times 32}$ ($H_{ij} \sim \N(0, 1/32)$), so the per-symbol sub-channel is identical for all $N$; BER improvements are primarily attributable to coding gain, though the code families (CCSDS and WiMax) differ.
We test five rate-$1/2$ LDPC codes: CCSDS $(128, 64)$, $(256, 128)$, $(512, 256)$ and WiMax $(1056, 528)$, $(2304, 1152)$.
Module~C exploits the block structure, requiring only a single $32 \times 32$ inversion per iteration.

\begin{figure}[t]
  \centering
  \includegraphics[width=\columnwidth]{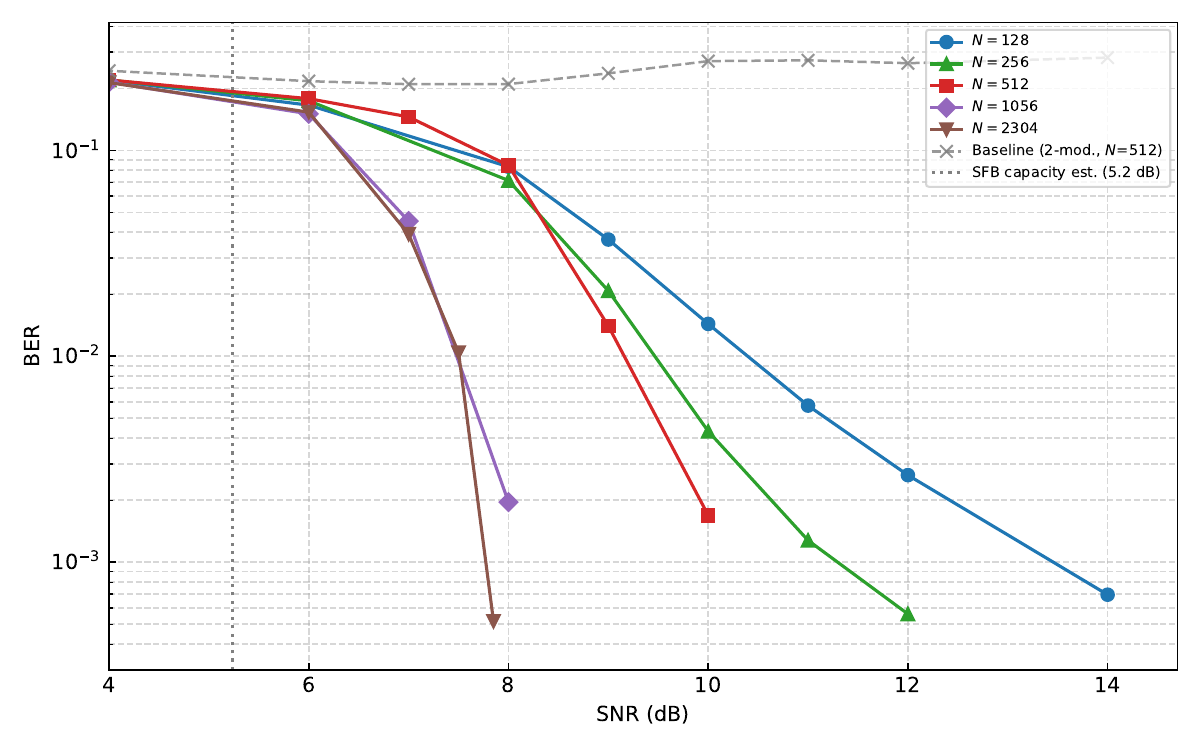}
  \caption{BER versus SNR for $\vy = \tanh(\mH\vx) + \vz$ with block-diagonal $\mH$ ($32 \times 32$ blocks).
  Solid curves: proposed 3-module SC-VAMP for $N = 128$--$2304$.
  Dashed gray: mismatched 2-module baseline ignoring the nonlinearity ($N = 512$).
  Dotted vertical line: SFB-based capacity estimate ($\approx 5.2$~dB)~\cite{wadayama2025sfb}.}
  \label{fig:ber_tanh}
\end{figure}

Fig.~\ref{fig:ber_tanh} shows the BER versus SNR for the five code lengths.
The proposed 3-module SC-VAMP exhibits a clear waterfall behavior for all $N$, with the waterfall slope steepening as the block length increases.
For the longest code ($N = 2304$), the BER reaches $\approx 5 \times 10^{-4}$ at $7.85$~dB and zero errors at $8$~dB (within 500 seeds).
The mismatched 2-module baseline ($N = 512$, dashed) remains at BER $\approx 0.2$--$0.3$ across all SNR, confirming that ignoring the nonlinearity prevents successful decoding.
The capacity estimate ($\approx 5.2$~dB) obtained via the SFB method~\cite{wadayama2025sfb} is shown as a dotted vertical line; the gap of $\sim$3--4~dB to the waterfall onset is consistent with short-to-moderate block length LDPC codes at rate~$1/2$.

\section{Conclusion}\label{sec:conc}

We proposed a 3-module SC-VAMP framework that separates a nonlinear channel into likelihood, coupling, and code-constraint modules, each equipped with score/Fisher-based Onsager-corrected extrinsic messaging.
The key advantage is modularity: Module~A with Gauss--Hermite quadrature handles any element-wise nonlinearity without modifying the other modules.
Experiments confirm that the Onsager correction is indispensable, yielding a $\sim$3~dB gain over variants without proper extrinsic exchange, and that the waterfall steepens consistently from $N = 128$ to $2304$.
Future directions include state evolution analysis, application to other nonlinearities (e.g., hard/soft clipping, 1-bit quantization), and data-driven score learning~\cite{wadayama2025scvamp} for unknown $f$.

\section*{Acknowledgment}
This work was supported by JST, CRONOS, Japan Grant Number JPMJCS25N5 and JSPS KAKENHI Grant Numbers JP25H01111 and JP26K00949.


\begin{thebibliography}{99}

\bibitem{donoho2009amp}
D.~L. Donoho, A.~Maleki, and A.~Montanari,
``Message-passing algorithms for compressed sensing,''
\emph{Proc. Natl. Acad. Sci.}, vol.~106, no.~45, pp.~18914--18919, 2009.

\bibitem{rangan2019vamp}
S.~Rangan, P.~Schniter, and A.~K. Fletcher,
``Vector approximate message passing,''
\emph{IEEE Trans. Inf. Theory}, vol.~65, no.~10, pp.~6664--6684, 2019.

\bibitem{ma2017oamp}
J.~Ma and L.~Ping,
``Orthogonal AMP,''
\emph{IEEE Access}, vol.~5, pp.~2020--2033, 2017.

\bibitem{bayati2011amp}
M.~Bayati and A.~Montanari,
``The dynamics of message passing on dense graphs, with applications to compressed sensing,''
\emph{IEEE Trans. Inf. Theory}, vol.~57, no.~2, pp.~764--785, 2011.

\bibitem{takeuchi2020vamp}
K.~Takeuchi,
``Rigorous dynamics of expectation-propagation-based signal recovery from unitarily invariant measurements,''
\emph{IEEE Trans. Inf. Theory}, vol.~66, no.~1, pp.~368--386, 2020.

\bibitem{guo2005mutual}
D.~Guo, S.~Shamai~(Shitz), and S.~Verd\'{u},
``Mutual information and minimum mean-square error in {Gaussian} channels,''
\emph{IEEE Trans. Inf. Theory}, vol.~51, no.~4, pp.~1261--1282, 2005.

\bibitem{rangan2011gamp}
S.~Rangan,
``Generalized approximate message passing for estimation with random linear mixing,''
in \emph{Proc. IEEE Int. Symp. Inf. Theory (ISIT)}, 2011, pp.~2168--2172.

\bibitem{schniter2016vamp}
P.~Schniter, S.~Rangan, and A.~K. Fletcher,
``Vector approximate message passing for the generalized linear model,''
in \emph{Proc. 50th Asilomar Conf. Signals, Systems, and Computers}, 2016, pp.~1525--1529.

\bibitem{minka2001ep}
T.~P. Minka,
``Expectation propagation for approximate {Bayesian} inference,''
in \emph{Proc. 17th Conf. Uncertainty in Artificial Intelligence (UAI)}, 2001, pp.~362--369.

\bibitem{opper2005ep}
M.~Opper and O.~Winther,
``Expectation consistent approximate inference,''
\emph{J. Mach. Learn. Res.}, vol.~6, pp.~2177--2204, 2005.

\bibitem{wadayama2025scvamp}
T.~Wadayama and T.~Takahashi,
``Score-based {VAMP} with {Fisher}-information-based {Onsager} correction,''
arXiv:2601.07095, 2026.

\bibitem{efron2011tweedie}
B.~Efron,
``Tweedie's formula and selection bias,''
\emph{J. Amer. Statist. Assoc.}, vol.~106, no.~496, pp.~1602--1614, 2011.

\bibitem{douillard1995turbo}
C.~Douillard \emph{et~al.},
``Iterative correction of intersymbol interference: Turbo-equalization,''
\emph{European Trans. Telecommun.}, vol.~6, no.~5, pp.~507--511, 1995.

\bibitem{tuchler2002turbo}
M.~T\"{u}chler, R.~Koetter, and A.~C. Singer,
``Turbo equalization: Principles and new results,''
\emph{IEEE Trans. Commun.}, vol.~50, no.~5, pp.~754--767, 2002.

\bibitem{wang1999iterative}
X.~Wang and H.~V. Poor,
``Iterative (turbo) soft interference cancellation and decoding for coded {CDMA},''
\emph{IEEE Trans. Commun.}, vol.~47, no.~7, pp.~1046--1061, 1999.

\bibitem{abramowitz1964handbook}
M.~Abramowitz and I.~A. Stegun,
\emph{Handbook of Mathematical Functions}.
Dover, 1964.

\bibitem{gallager1962ldpc}
R.~G. Gallager,
``Low-density parity-check codes,''
\emph{IRE Trans. Inf. Theory}, vol.~8, no.~1, pp.~21--28, 1962.

\bibitem{richardson2008modern}
T.~J. Richardson and R.~L. Urbanke,
\emph{Modern Coding Theory}.
Cambridge Univ. Press, 2008.

\bibitem{wadayama2025sfb}
T.~Wadayama,
``Mutual information estimation via score-to-{Fisher} bridge for nonlinear {Gaussian} noise channels,''
arXiv:2510.05496, 2025.

\bibitem{kaira2025}
{Kaira Contributors},
``Kaira: A {PyTorch}-based toolkit for simulating communication systems,''
2025. [Online]. Available: \url{https://github.com/ipc-lab/kaira}

\end{thebibliography}
\end{document}